\documentclass[conference]{IEEEtran}
\IEEEoverridecommandlockouts
\usepackage{algorithm}
\usepackage{float} 
\usepackage{cite}
\usepackage{url} 
\usepackage{amsmath,amssymb,amsfonts}
\usepackage{graphicx}
\usepackage{textcomp}
\usepackage{xcolor}
\def\BibTeX{{\rm B\kern-.05em{\sc i\kern-.025em b}\kern-.08em
    T\kern-.1667em\lower.7ex\hbox{E}\kern-.125emX}}

\makeatletter
\let\old@ps@headings\ps@headings
\let\old@ps@IEEEtitlepagestyle\ps@IEEEtitlepagestyle
\def\confheader#1{%
\def\ps@IEEEtitlepagestyle{
\old@ps@IEEEtitlepagestyle
\def\@oddhead{\strut\hfill#1\hfill\strut}
\def\@evenhead{\strut\hfill#1\hfill\strut}
}
\ps@headings
}
\makeatother
\confheader{
\small{Proceedings of the 13th RSI International Conference on Robotics and Mechatronics (ICRoM 2025), Dec. 16-18, 2025, Tehran, Iran} 
}
\usepackage[pscoord]{eso-pic}
\newcommand{\placetextbox}[3]{
\setbox0=\hbox{#3}
\AddToShipoutPictureFG*{ \put(\LenToUnit{#1\paperwidth},\LenToUnit{#2\paperheight}){\vtop{{\null}\makebox[0pt][c]{#3}}}
}
}
\placetextbox{.5}{0.055}{\textbf{\small{Proceedings of the 13th RSI International Conference on Robotics and Mechatronics (ICRoM 2025), Dec. 16-18, 2025, Tehran, Iran}}}

\begin{document}

\title{Optimal Regulation of Nonlinear Input-Affine Systems via an Integral Reinforcement Learning–Based State-Dependent Riccati Equation Approach
}
\author{
\IEEEauthorblockN{Arya Rashidinejad Meibodi}
\IEEEauthorblockA{
Advanced Service Robots (ASR) \\
Laboratory, Department of \\
Mechatronics Engineering, School \\
of Intelligent Systems Engineering,\\
College of Interdisciplinary Science \\
and Technology, University of Tehran\\
Tehran, Iran\\
aria.rashidi.nm@alumni.ut.ac.ir
}

\and
\IEEEauthorblockN{ Mahbod Gholamali Sinaki}
\IEEEauthorblockA{Department of Mechanical Engineering \\
\textit{K. N. Toosi University of Technology}\\
Tehran, Iran \\
sinaki@email.kntu.ac.ir}
\and
\IEEEauthorblockN{Khalil Alipour}
\IEEEauthorblockA{
Advanced Service Robots (ASR) \\
Laboratory, Department of \\
Mechatronics Engineering, School \\
of Intelligent Systems Engineering,\\
College of Interdisciplinary Science \\
and Technology, University of Tehran\\
Tehran, Iran\\
k.alipour@ut.ac.ir
}
}

\maketitle
\begin{abstract}
The State-Dependent Riccati Equation (SDRE) technique generalizes the classical algebraic Riccati formulation to nonlinear systems by designing an input to the system that optimally (suboptimally) regulates system states toward the origin while simultaneously optimizing a quadratic performance index. In the SDRE technique, we solve the State-Dependent Riccati Equation to determine the control for regulating a nonlinear input-affine system. Since an analytic solution to SDRE is not straightforward, one method is to linearize the system at every state, solve the corresponding Algebraic Riccati Equation (ARE), and apply optimal control until the next state of the system. Completing this task with high frequency gives a result like the original SDRE technique. Both approaches require a complete model; therefore, here we propose a method that solves ARE in every state of the system using a partially model-free approach that learns optimal control in every state of the system, without explicit knowledge of the drift dynamics, based on Integral Reinforcement Learning (IRL). To show the effectiveness of our proposed approach, we apply it to the second-order nonlinear system in simulation and compare its performance with the classical SDRE method, which relies on the system's model and solves the ARE at each state. Our simulation results demonstrate that, with sufficient iterations, the IRL-based approach achieves approximately the same performance as the conventional SDRE method, demonstrating its capability as a reliable alternative for nonlinear system control that does not require an explicit environmental model.
\end{abstract}


\section{Introduction}
\label{sec:Introduction}

Regulation control is a crucial problem in control, focused on bringing the system states to a fixed chosen reference.
In linear systems, the system states can be driven to the origin by shifting the eigenvalues to the left side of the imaginary axis of the complex plane, ensuring that the norm of the state vector approaches zero asymptotically. This can be implemented through a feedback mechanism that utilizes the system's state variables, incorporating them into the control input to adjust the placement of eigenvalues in the closed-loop system. This is an idea behind the state feedback controller \cite{kalman1960contributions}. For regulation to other points rather than the origin, we can first transform the states of the system, then do the regulation process through the origin. Optimal regulation is another problem that addresses this problem from an optimization perspective; in other words, we should calculate an input that minimizes a performance criterion while adhering to the system's model. For the minimization of the cost function in the infinite time horizon, states must be regulated toward the desired points. 
Such regulation is obtained by computing the control input that minimizes a defined performance criterion \cite{mayne2000constrained}.
In cases where the cost function is not quadratic or includes constraints other than dynamics, the optimization problem should be solved directly. Given a convex problem and constraints, this process can be accomplished using convex optimization techniques \cite{boyd2004convex}.

For the problem where we are only constrained by the dynamics and have a quadratic cost function, determining the optimal feedback gain requires solving ARE \cite{mehrmann1991autonomous}.
When our cost function is a quadratic function and is defined over an infinite time horizon, solving the ARE is a straightforward approach that yields the best result \cite{willems2003least}, and when defined over a limited time horizon, the optimization problem can be resolved through the differential Riccati equation, which can be solved numerically backward in time \cite {kalman1960contributions}.
The applicability of these methods is confined to linear dynamic systems. The nonlinear case demands the Hamilton-Jacobi-Bellman (HJB) equation as a key to optimality \cite{lewis2012optimal}. However, because of the complexity of generalizing the HJB equation method, in this article, we use the SDRE technique, which is a generalization of ARE for the nonlinear input affine case \cite{cloutier1996nonlinear}, \cite{cloutier1996nonlinearpart2}, \cite{ccimen2008state}. The SDRE technique has been extensively used for multiple real-world problems \cite{do2011sdre}, \cite{razzaghi2019sliding}. 
Massari et al. \cite{massari2014application} applied the SDRE framework to spacecraft attitude and orbital regulation. In \cite{strano2015sdre}, SDRE has been used for hydraulic actuation systems. In the SDRE technique, we begin by converting the nonlinear system into a State-Dependent Coefficient (SDC) format. Afterward, by solving the related Riccati equation whose coefficients are not constant, we achieve an optimal (or suboptimal, in the multistate case) solution. The suboptimality in the multi-state case is because the parametrization process is not unique in the multivariable case \cite{ccimen2008state}, \cite{cloutier1996nonlinear}. This illustrates the importance of the parametrization manner in the SDRE methodology, and \cite{liang2013analysis} investigates the problem of finding the appropriate SDC format. The analytical solution to SDRE is complex; therefore, in most cases, we can linearize the system at every state, solve the corresponding ARE, and employ the derived control input. This process achieves the same result as the original SDRE method when performed at a high frequency \cite{cloutier1996nonlinear},\cite{da2021fast}.
IRL is another approach that can be used to iteratively achieve a solution to ARE. This iterative solution was first discussed in \cite{kleinman1968iterative}. Following this, in \cite{vrabie2009adaptive} and \cite{jiang2012computational}, partially model-free and completely model-free versions were designed, respectively. In this article, we introduce a novel IRL-based SDRE approach, which, instead of solving the ARE in every state, obtains a solution by applying a sufficient number of IRL iterations.
The use of IRL to solve SDRE was used in \cite{batmani2016state}. They used the IRL algorithm stated in \cite{vrabie2009neural} to directly regulate a state-dependent system, which is a nonlinear system. Unlike \cite{batmani2016state}, which employed nonlinear IRL for solving nonlinear state-dependent dynamics, we combine the conventional SDRE method with the linear IRL method. Compared with the approach in \cite{batmani2016state}, which suffers from slow convergence due to the nonlinear nature of the value function requiring neural approximation, the present method demonstrates superior learning speed. Aside from the partially model-free nature of IRL, which will be beneficial for our method, IRL can be used with non-quadratic cost functions, which is a point that can be beneficial in some control cases, and also using IRL for solving ARE removes dependency on linear algebraic toolboxes in MATLAB or Python.

The rest of the paper reviews previous theoretical concepts and our proposed method, and finally presents the experimental evaluation. In section~\ref{sec:SDRE}, the SDRE technique will be discussed completely. Section~\ref{sec:IRL} investigates the possibility of solving the ARE employing the Integral Reinforcement Learning approach by repeatedly alternating between value estimation (through integration) and improvement of policy. In Section~\ref{sec:IRLSDRE}, the SDRE technique will be extended, and Integral Reinforcement Learning will be incorporated in the classical method. Section~\ref{sec:Simulation}, reports results of the simulation, and the results of the IRL-based SDRE will be evaluated against the classical SDRE method. Section~\ref{sec:conclusion} outlines the main findings along with suggestions for future investigations.

\section{State Dependent Riccati Equation Technique}
\label{sec:SDRE}

In this part, the SDRE framework is analyzed, and its main features are outlined in detail. SDRE is a powerful method for achieving optimal regulation of nonlinear input-affine systems based on a quadratic performance criterion, which is the core of this paper. The SDRE technique begins by parameterizing the nonlinear system, in other words, transforming it into a linear-like structure where the system matrices are state-dependent. Parametrization provides us with the drift and input matrices, which are smooth functions of the state and can be found in the SDRE formulation. Solution to SDRE yields the optimal (or suboptimal in multi-state cases) control law for regulating the nonlinear system. By solving the SDRE, the approach generates a feedback controller that optimizes the quadratic performance measure, offering near-optimal performance for a broad class of nonlinear dynamics.

Since analytical solutions to the SDRE are rarely feasible due to their matrices changing with the current state of the system, a practical alternative involves linearizing the system at the current state, finding the corresponding ARE solution, and then acquiring optimal input, which remains active until the next state is reached. When this linearization and ARE solution-finding process is performed at a sufficiently high frequency, the resulting control closely approximates the original SDRE solution. The subsequent subsections outline this process. The first subsection discusses the general regulation problem for systems with nonlinear dynamics. The second subsection describes system parameterization and the reasons why the SDRE technique approaches suboptimal solutions in problems with more than one state variable. The third presents the SDRE algorithm, and the fourth addresses implementation limitations along with the conventional high-frequency ARE-solving approach.

\subsection{Nonlinear Regulation Problem}
\label{subsec:NonlinearRegulation}

The nonlinear regulation problem refers to the task of regulating the states of a nonlinear system to specific points. In this article, when we refer to nonlinear regulation, we aim to solve an optimization problem that yields a near-optimal control, minimizing the cost function subject to the nonlinear input-affine dynamics. This problem is called the optimal regulation of a nonlinear input-affine system. In most cases, a quadratic function of the state vector and the input vector is typically used as the cost function.

Assume a nonlinear system whose dynamics are affine with respect to the control input, formulated as follows:
\begin{equation}
\dot{x} = f(x) + g(x)u
\end{equation}
In this formulation, $x\in\mathbb{R}^n$ denotes the system state vector and $u\in\mathbb{R}^m$ the control input. The mappings $f(x)$ (drift) and $g(x)$ characterize the dynamics and are assumed to be smooth functions. We need to find optimal \(u(x)\) that optimizes the performance criterion represented below:
\begin{equation}
J(x_0, u) = \int_{t_0}^\infty \left( x^T Q x + u^T R u \right) dt
\end{equation}

subject to the system dynamics, where weighting matrices \(Q \) and \(R \) should be positive semi-definite and positive definite, respectively, ensuring the cost penalizes deviations from the equilibrium (typically the origin) and control effort.

The Linear Quadratic Regulator (LQR) problem is identical; however, it is constrained to linear dynamics. This formulation is known as the Nonlinear Quadratic Regulator (NQR) problem.
A common method for tackling this problem is to derive the HJB equation and determine its solution, which defines the criterion that an optimal solution must satisfy. Assuming a value function \(V(x)\) representing the minimum cost from state \(x\), the stationary HJB equation is:
\begin{equation}
\min_u \left\{ x^T Q x + u^T R u + \nabla V^T \left( f(x) + g(x)u \right) \right\} = 0
\label{eq:HJB}
\end{equation}

By differentiating with respect to $u$ and setting to zero, the formulation for the optimal input  will be: 
\begin{equation}
u^* = -\frac{1}{2} R^{-1} g(x)^T \nabla V
\label{eq:diff}
\end{equation}

Substituting $u^*(x)$ back into~\eqref{eq:HJB} gives a nonlinear partial differential equation (PDE) in $V(x)$. Solving this PDE gives $\nabla V(x)$, and substituting it into~\eqref{eq:diff} produces the optimal control input; however, because solving this PDE is intractable for most nonlinear cases, approximate methods like the SDRE have broad applicability in real engineering problems. Although an analytical solution to SDRE remains complex, as with HJB, there are approximate methods, such as conventional SDRE, that, instead of solving SDRE analytically, solve ARE at every system state and can replace the traditional SDRE method. Because of this, expanding SDRE can be beneficial in lots of real-world applications.

The SDRE method can be seen as a general form of ARE used in LQR, delivering an efficient approximation to the infinite-horizon regulation problem in nonlinear control. In the SDRE method, by parameterizing the nonlinear input affine dynamics, a state-dependent linear-like form is obtained. After parametrization, by solving the SDRE, which is like the ARE with state-dependent coefficients, a suboptimal but effective feedback control is derived. 

\subsection{System Parameterization}
\label{subsec:Parametrization}

The initial stage in applying the SDRE technique is to parameterize the nonlinear input-affine system and transform it into a form known as the SDC form. This process involves rewriting the nonlinear input-affine dynamics \(\dot{x} = f(x) + g(x) u\) as a linear-like structure with state-dependent matrices, specifically \(f(x) = A(x) x\) and \(g(x) = B(x)\), where \(A(x) \in \mathbb{R}^{n \times n}\) and \(B(x) \in \mathbb{R}^{n \times m}\) are functions of the state \(x\). Under SDC parametrization, the system is expressed in a form resembling a time-invariant linear model, but its matrices are functions of the system state. SDRE, which will be discussed later, is a generalization of the ARE for this nonlinear system with an SDC form.
The parameterization process is not unique for systems with more than one state variable (\(n > 1\)), as multiple valid decompositions of \(f(x)\) into \(A(x) x\) may exist \cite{ccimen2008state,cloutier1996nonlinear}. In contrast, for scalar systems (\(n = 1\)), the parameterization is unique \cite{ccimen2008state,cloutier1996nonlinear}. This non-uniqueness in multi-state cases implies that the SDRE technique yields the optimal solution only for scalar systems; for multi-state systems, it produces a near-optimal control strategy whose performance depends on the type of parameterization. 

Under mild assumptions, a smooth parametrization, and stabilizable and detectable parametrization in a region around a desired point, the SDRE approach has an asymptotic stability proof \cite{cloutier1996nonlinear}.

\subsection{SDRE Algorithm}
The SDRE framework offers a structured method for deriving an optimal (or suboptimal, in the multi-state case) control law for systems with nonlinear input-affine dynamics.

The algorithm begins with the transformation of system dynamics \(\dot{x} = f(x) + g(x)u\) into a SDC form, where \(f(x) = A(x)x\) and \(g(x) = B(x)\). Next, an SDRE is solved to acquire the positive definite matrix \(P(x)\), which is used to compute the optimal feedback gain. The resulting control is applied to regulate the system optimally (or suboptimally for multi-state systems) with respect to the consequent formulation, which corresponds to an optimal control task defined over an infinite horizon.
\begin{equation}
PI\footnote{Performance Index} = \int_{t_0}^\infty (x^T Q x + u^T R u) dt
\end{equation}

Algorithm~\ref{alg:sdre} shows the general form of the SDRE algorithm.

\begin{algorithm}[H]
  \caption{SDRE Algorithm (theoretical ideal)}
  \label{alg:sdre}
    \textbf{Dynamics and Performance Index:}
  \[
    \dot{x} = f(x) + g(x)u \qquad
    PI = \int_{t_0}^\infty \!\left( u^\top R(x) u + x^\top Q(x) x \right) dt
  \]
   \textbf{Parametrization:} Represent $f(x) = A(x)x$, $g(x) = B(x)$,
    where the pair $(A(x),B(x))$ is stabilizable and $(Q^{1/2}(x),A(x))$ is detectable
    in a neighborhood of the origin.
   \textbf{Solution to SDRE:} Solve the provided equation for the positive definite function $P(x)$:
  \[
    A(x)^\top P(x) + P(x)A(x) - P(x)B(x)R^{-1}(x)B(x)^\top P(x)
  \]
  \[
    = -Q(x)
  \]
   \textbf{Control law:} Compute the optimal input
  \[
    u(x) = -\,R(x)^{-1} B(x)^\top P(x)\,x
  \]
\end{algorithm}

This optimal feedback controller law enables the real-time application of the SDRE technique, which effectively regulates the nonlinear system state.

\subsection{Limitations of SDRE Technique}
\label{subsec:Limitation}

The SDRE framework, while effective for nonlinear optimal regulation, has a notable limitation due to the complex analytical solution to the SDRE. 
Unlike the ARE used for the LQR problem, the SDRE's state-dependent structure yields a nonlinear and coupled set of equations that are difficult to solve in closed form. Analytical solutions are rarely feasible, except in highly simplified cases. To address this, a conventional approximation algorithm is commonly employed. 

The conventional SDRE Algorithm involves linearizing the system at every state, solving the corresponding ARE for that linearized model, and applying the resulting optimal control until the next state in simulation. The process then repeats with relinearization in the updated state. When performed at a sufficiently high frequency, this approach yields results that approximate the performance of the standard SDRE method, without requiring direct calculation of the SDRE solution.
The conventional SDRE Algorithm is shown in Algorithm~\ref{alg:ConvSDRE}.

\begin{algorithm}
\caption{Conventional SDRE Algorithm (Practical version)}
\label{alg:ConvSDRE}
 \textbf{Dynamics and Performance Index:} \[
        \dot{x} = f(x) + g(x)u
        \qquad 
        PI = \int_{t_0}^\infty \!\left(u^\top R(x) u + x^\top Q(x) x \right) dt
      \]
    
 \textbf{Parametrization:} Represent $f(x) = A(x)x$, $g(x) = B(x)$, 
           where the pair $(A(x),B(x))$ is stabilizable and $(Q^{1/2}(x),A(x))$ is detectable 
           in a neighborhood of the origin.
    
 \textbf{Procedure:}     
    At each time step:
    \begin{enumerate}
        \item Linearize the system at the current state $x(t)$ to obtain constant matrices $A = A(x(t))$, $B = B(x(t))$, $Q = Q(x(t))$, and $R = R(x(t))$.
        \item Solve the Algebraic Riccati Equation (ARE): 
        \[
        A^T P + P A - P B R^{-1} B^T P + Q = 0
        \]
        for a positive definite matrix $P \succ 0$.
        \item Apply the control $u(x)$ until the next time step. 
        \[
        u(x) = -R(x)^{-1} B(x)^T P x.
        \]
        
        \item Relinearize at the new state and repeat the procedure.
    \end{enumerate}
\end{algorithm}

\section{Integral Reinforcement Learning for LQR Problems}
\label{sec:IRL}

The study presented in \cite{kiumarsi2017optimal} demonstrates the use of reinforcement learning (RL) for tackling optimal control tasks in continuous and discrete-time settings. IRL and Inverse Reinforcement Learning (Inverse RL) are two main RL methodologies that have been extensively applied in control theory. IRL is a type of RL approach, mostly used for continuous-time systems. Vrabie et al. introduced the partially model-free IRL for continuous-time LQR problems \cite{vrabie2009adaptive}. This approach learns the optimal input (without knowing the drift dynamics) by running multiple iterations of Integral Reinforcement Learning (IRL). Using an Iterative approach to find optimal control of the regulation problem is not a new development and was first implemented in \cite{kleinman1968iterative}, however, In \cite{vrabie2009adaptive}, the authors developed an approach that operates without complete model knowledge that iteratively approximates the optimal control law by converging to the positive definite matrix satisfying the ARE without full model knowledge. More specifically, for detectable and stabilizable linear systems, IRL starts from an initial stabilizing policy and iteratively applies control to the system, integrates the cost, and improves the policy. This IRL algorithm is shown in Algorithm~\ref{alg:IRL}. The IRL method implements online policy iteration using the integral form of the continuous-time Bellman equation:

\begin{equation}
\begin{aligned}
& x(t)^\top P_i x(t) - x(t+\delta)^\top P_i x(t+\delta) \\
&\quad = \int_{t}^{t+\delta} \left( x(\tau)^\top Q x(\tau) + u(\tau)^\top R u(\tau) \right) d\tau
\end{aligned}
\end{equation}
which eliminates the need for the unknown drift matrix $A$.
During each iteration $i$, a noise signal $\xi(t)$ is added to the current policy:
 \begin{equation}
     u_i(t) = -K_i x(t) +\xi(t)
 \end{equation} to ensure persistent excitation. Usually a sum of sinusoids is employed:
\begin{equation}
    \xi(t) = \sum_{k=1}^{N_f} a_k \sin(\omega_k t)
\end{equation}

although small zero-mean Gaussian noise is sometimes used in purely numerical experiments.
The quadratic value function $V_i(x) = x^\top P_i x$ is then estimated by solving a least-squares problem constructed from multiple time intervals, followed by policy improvement via $K_{i+1} = R^{-1} B^\top P_i$. This data-driven alternation between policy evaluation and improvement converges to the optimal solution of the Algebraic Riccati Equation without ever forming or solving the ARE explicitly, making the approach particularly attractive for subsequent extension to state-dependent Riccati equation (SDRE) problems. In the same year another version was introduced for the nonlinear regulator problem \cite{vrabie2009neural}.

\begin{algorithm}
  \caption{Integral Reinforcement Learning for LQR}
  \label{alg:IRL}
     \textbf{Dynamics and Performance Index:} (Assume $A$ unknown; $B$, $Q$, $R$ known)
      \[
        \dot{x} = A x + B u 
        \qquad 
        PI = \int_{0}^{\infty} \left( u^\top R u +x^\top Q x\right) dt
      \]
      \[
        Q \succeq 0,\; R \succ 0.
      \]
     \textbf{Initialization:} Initialize the process using a control law that stabilizes the system
    \[u = -K_0 x\]
     \textbf{Procedure:} For iterations $i = 0,1,\dots,N-1$:
     \quad \begin{enumerate}
      \item \textit{Data collection:} Apply control
        \[
          u(t) = -K_i x(t) + \xi(t),
        \]
        where $\xi(t)$ is a small exploration signal. Collect state $x(t)$ and input $u(t)$ at time points $t_1, t_2, \dots, t_M$ over a time window $[t_0, t_f]$.
      \item \textit{Policy evaluation:} Estimate the value function $V_i(x) = x^\top P_i x$ ($P_i \succ 0$) for the policy $u = -K_i x$. For $M$ small intervals $[t_k, t_k + \delta]$, compute numerically:
      \[
  \int_{t_k}^{t_k + \delta} \big[\, u(t)^\top R u(t) + x(t)^\top Q x(t) \,\big]\, dt
\]

        Form the equation for each interval:
        \[
          x(t_k)^\top P_i x(t_k) - x(t_k + \delta)^\top P_i x(t_k + \delta) = \text{[Integral]}.
        \]
        Stack equations for $k=1,\dots,M-1$ and solve for $P_i$ using least-squares.
      \item \textit{Policy improvement:} Update the control gain:
        \[
          K_{i+1} = R^{-1} B^\top P_i
        \]
    \end{enumerate}
     \textbf{Output:} Converged control gain $K^* \approx K_N$ and control law \[u = -K^* x\]
\end{algorithm}

In this work, instead of solving ARE, we employ algorithm~\ref{alg:IRL} to approximate the ARE solution for optimal control of linear systems. The completely model-free approach used in \cite{jiang2012computational} can also be employed as an alternative to Algorithm~\ref{alg:IRL}. 

Algorithm~\ref{alg:IRL} offers key advantages, first advantage is its applicability to non-quadratic cost functions, and more important benefit of this method lies in its semi–model–free structure, which allows it to learn directly through system interactions rather than relying on an explicit environmental model.

\section{IRL-Based SDRE}
\label{sec:IRLSDRE}

An enhanced version of the SDRE framework, incorporating IRL, is outlined in this section. The enhancement is accomplished by replacing the direct algebraic ARE solution with IRL iterations. Traditional implementations (Conventional SDRE method in Algorithm~\ref{alg:ConvSDRE}) rely on linear algebraic toolboxes in MATLAB or Python to solve the ARE at each state.

The IRL-based SDRE method, as expanded in this section, incorporates partially model-free IRL into the conventional SDRE method. This means that at every state, after parameterizing or linearizing the system, we initialize with a stabilizing control gain and perform sufficient IRL iterations to converge to the optimal gain corresponding to the local ARE solution.

This integration enables online learning and adaptation without requiring explicit knowledge of the full model or computational solvers, making it suitable for real-time applications where system dynamics may be partially unknown. More importantly, if we start from a stabilizing initial gain, running sufficient IRL iterations ensures that we have the same control gain as by designing a gain by the ARE solution. This means that by applying sufficient IRL iterations, this approach converges to the conventional SDRE. The IRL-based SDRE methodology is illustrated in Algorithm~\ref{alg:IRL_SDRE}.

\begin{algorithm}
  \caption{IRL-based SDRE Algorithm}
  \label{alg:IRL_SDRE}
     \textbf{Dynamics and Performance Index:} 
      \[
        \dot{x} = f(x) + g(x)u
        \qquad 
        PI = \int_{t_0}^\infty \left(u^\top R(x) u + x^\top Q(x) x\right) dt
      \]
     \textbf{Parametrization:} Represent $f(x) = A(x)x$, $g(x) = B(x)$, 
           where the pair $(A(x), B(x))$ is stabilizable and $(Q(x)^{1/2}, A(x))$ is detectable 
           in a neighborhood of the origin. Assume $A(x)$ unknown; $B(x)$, $Q(x)$, $R(x)$ known.
     \textbf{Procedure:} At each time step $t$:
     \quad \begin{enumerate}
      \item \textit{Linearization:} Linearize the system at the current state $x(t)$ to obtain constant matrices $A = A(x(t))$, $B = B(x(t))$, $Q = Q(x(t))$, and $R = R(x(t))$.
      \item \textit{IRL Iterations:} Initialize a stabilizing control law $u = -K_0 x$. For iterations $i = 0,1,\dots,N-1$:
        \begin{enumerate}
          \item \textit{Data collection:} Apply control
        \[
          u(t) = -K_i x(t) + \xi(t),
        \]
        where $\xi(t)$ is a small exploration signal. Collect state $x(t)$ and input $u(t)$ at time points $t_1, t_2, \dots, t_M$ over a time window $[t_0, t_f]$.
      \item \textit{Policy evaluation:} Estimate the value function $V_i(x) = x^\top P_i x$ ($P_i \succ 0$) for the policy $u = -K_i x$. For $M$ small intervals $[t_k, t_k + \delta]$, compute numerically:
        \[
  \int_{t_k}^{t_k + \delta} \big[\, u(t)^\top R u(t) + x(t)^\top Q x(t) \,\big]\, dt
\]
        Form the equation for each interval:
        \[
          x(t_k)^\top P_i x(t_k) - x(t_k + \delta)^\top P_i x(t_k + \delta) = \text{[integral]}
        \]
        Stack equations for $k=1,\dots,M-1$ and solve for $P_i$ using least-squares.
      \item \textit{Policy improvement:} Update the control gain:
        \[
          K_{i+1} = R^{-1} B^\top P_i
        \]
        \end{enumerate}
      \item \textit{Apply control:} Use the converged or final $P_N$ and apply control until next state:
        \[
          u(t) = -R(x(t))^{-1} B(x(t))^\top P_N x(t).
        \]
      \item \textit{Relinearize:} Move to the next time step, update $x(t)$, and repeat.
    \end{enumerate}
\end{algorithm}

\section{Results and Discussion}
\label{sec:Simulation}

A comparative analysis between the SDRE and IRL–based SDRE methodologies is carried out in this study using a nonlinear model adapted from \cite{cloutier1996nonlinearpart2}. We aim to regulate its states toward origin.

The nonlinear system used in this section can be described by the following nonlinear dynamics:
\begin{equation}
    \dot{x}_1 = x_1 - x_1^3 + x_2 + u
\end{equation}
\begin{equation}
    \dot{x}_2 = x_1 + x_1^2 x_2 - x_2 + u
\end{equation}

We can write this dynamics in the form of an input-affine nonlinear setting:

\begin{equation}
    \dot{x} = f(x) + B u
\end{equation}

where:

\begin{equation}
f(x) = \begin{bmatrix} x_1 - x_1^3 + x_2 \\[4pt] x_1 + x_1^2 x_2 - x_2 \end{bmatrix}, \quad 
B = \begin{bmatrix} 1 \\[4pt] 1 \end{bmatrix}
\end{equation}

To apply the SDRE and IRL-based SDRE, first, we need to transform the nonlinear setting into SDC format; in other words, we parametrize the system as:

\begin{equation}
    \dot{x} = A(x) x + B u
\end{equation}  

using the below \( A(x) \) and \( B \).

\begin{equation}
A(x) = \begin{bmatrix} 1 - x_1^2 & 1 \\[4pt] 1 + x_1 x_2 & -1 \end{bmatrix}, \quad 
B = \begin{bmatrix} 1 \\[4pt] 1 \end{bmatrix}
\end{equation}

The quadratic performance criterion has hyperparameters which are set as \( Q = I_{2 \times 2} \) (state weighting matrix) and \( R = 1 \) (input weighting scalar).

We initialized all simulations from the point \( x_0 = [3, 1]^T \) over a time span of \( [0, 10] \) seconds.

This smooth parametrization ensures that the pair \( (A(x), B) \) is stabilizable and \( (A(x), \sqrt{Q}) \) is detectable around the origin in a suitable region, enabling the use of the SDRE. At the origin, \( A(0) = \begin{bmatrix} 1 & 1 \\ 1 & -1 \end{bmatrix} \), and the controllability matrix for \( (A(0), B) \) has determinant \( -1 \neq 0 \), confirming controllability (hence stabilizability). Since \( Q = I_{2 \times 2} > 0 \), detectability is satisfied. Also, this parametrization is smooth, which is mandatory for the SDRE method.

\subsection{Conventional SDRE Controller (practicl algorithm)}

First, we applied the conventional SDRE method, obtaining the ARE solution point-wise in each state using MATLAB's \texttt{care} function to compute state-dependent feedback gains \( K(x) = R^{-1} B^T P \).
The main closed-loop simulation horizon is \(T_{\mathrm{sim}}=10~\text{s}\) (ode45 over \([0,\,T_{\mathrm{sim}}]\)), with the performance cost accumulated via the trapezoidal rule on \([0,\,T_{\mathrm{sim}}]\).

\begin{figure}[ht]
    \centering
    \includegraphics[width=0.5\textwidth]{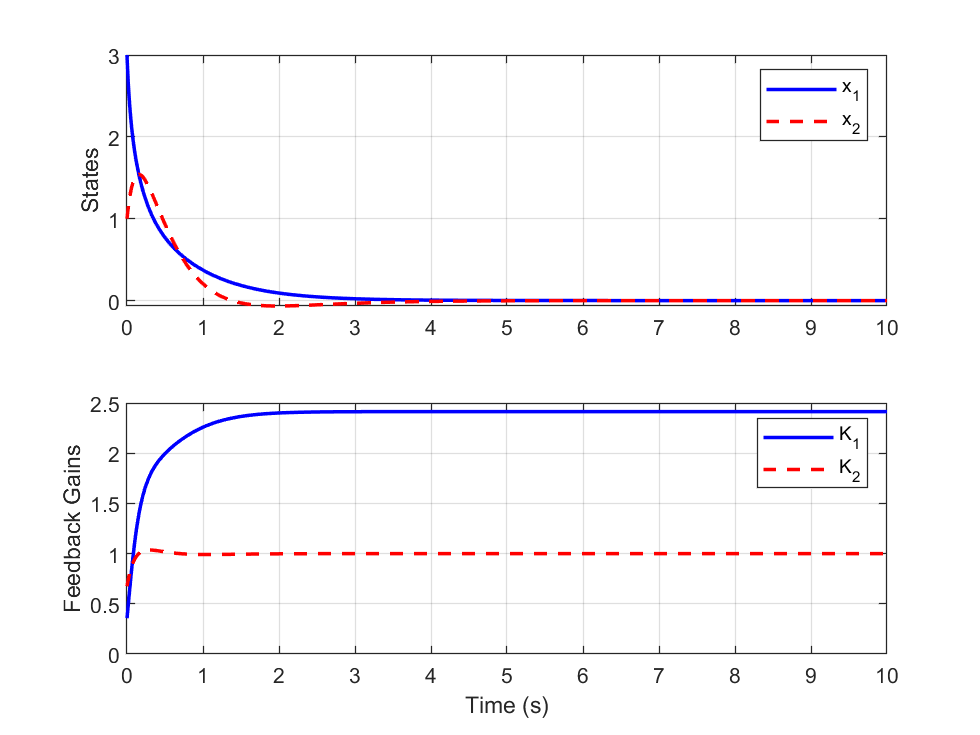}
    \caption{State trajectories (\( x_1 \), \( x_2 \)) and feedback gains (\( K_1 \), \( K_2 \)) for the conventional SDRE controller on the nonlinear system.}
    \label{fig:sdre_new}
\end{figure}

Figure~\ref{fig:sdre_new} shows the state trajectories and feedback gains (\( K_1 \), \( K_2 \)) over time, demonstrating effective regulation to the origin.

\subsection{IRL-based SDRE Controller}

\begin{figure}[ht]
    \centering
    \includegraphics[width=0.5\textwidth]{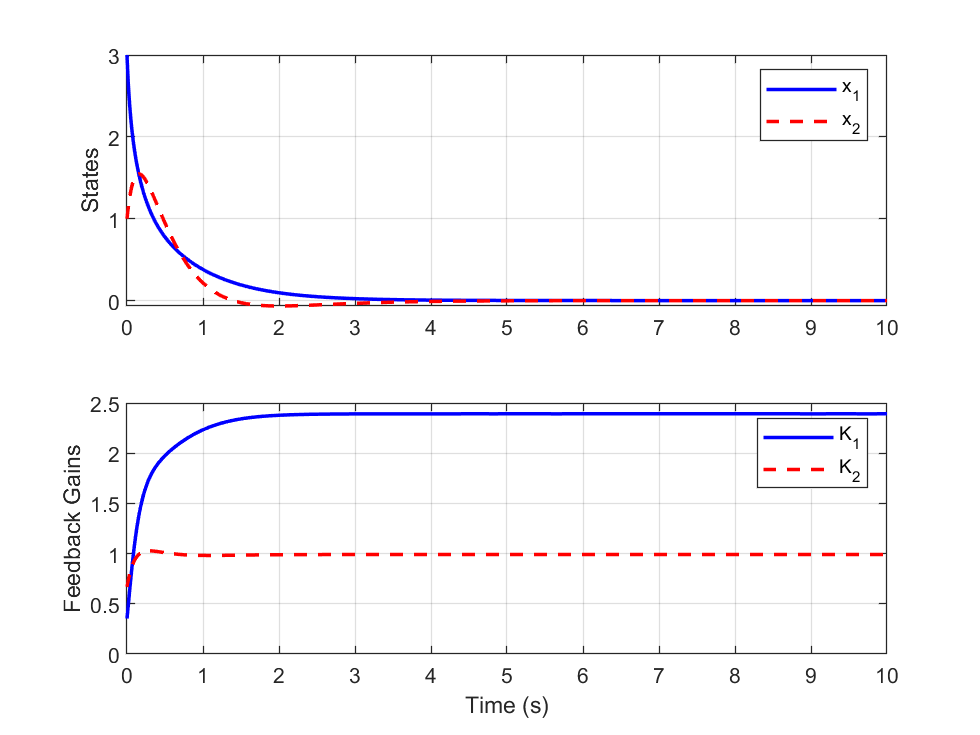}
    \caption{State trajectories (\( x_1 \), \( x_2 \)) and feedback gains (\( K_1 \), \( K_2 \)) for the IRL-based SDRE controller on the nonlinear system.}
    \label{fig:irlsdre_new}
\end{figure}

Next, we implemented an IRL-based SDRE controller, approximating the ARE solution at each state using IRL, a partially model-free approach that learns optimal control in every state of the system.
In this simulation we used the following hyperparameters: a maximum of \(N=100\) policy iterations; \(M=15\) sampled trajectories per iteration; sampling period \(T_s=0.001~\text{s}\); integration horizon \(\delta=0.1~\text{s}\); additive Gaussian exploration noise of magnitude \(0.01\) injected into the control input \(u\) to ensure persistent excitation; and a convergence tolerance \(\|K_{j+1}-K_j\|<10^{-4}\). The initial stabilizing feedback is obtained via pole placement with poles \(\{-1,-2\}\), i.e., \(K_0=\mathrm{place}(A,B,[-1,-2])\).
Figure~\ref{fig:irlsdre_new} illustrates the state trajectories and feedback gains, showing comparable regulation performance to the conventional SDRE, with a slight difference in performance. 

In the final simulation, we used 100 iterations per state and achieved results approximately equivalent to those of the conventional SDRE method. If we aim to get the same result as conventional, we should increase the number of maximum iterations to achieve the convergence tolerance of \( 10^{-4} \) or even less. The desired convergence tolerance and maximum number of iterations can be used, but strict limits require more computation time. It should be noted that most of the computational burden of the IRL algorithm lies in the least-squares step used in policy evaluation.

\subsection{Uncontrolled System}

\begin{figure}[ht]
    \centering
    \includegraphics[width=0.5\textwidth]{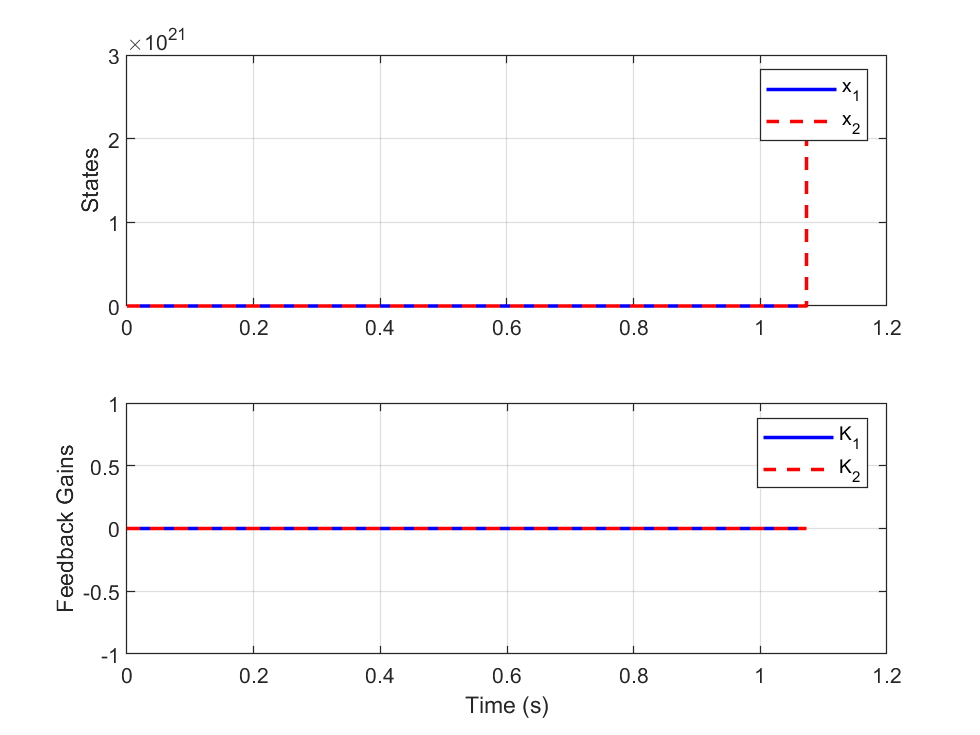}
    \caption{State trajectories (\( x_1 \), \( x_2 \)) and zero feedback gains (\( K_1 \), \( K_2 \)) for the uncontrolled nonlinear system.}
    \label{fig:uncontrolled_new}
\end{figure}

We also simulated the open-loop (uncontrolled) system, where \( u = 0 \), resulting in dynamics \( \dot{x} = A(x) x \). Figure~\ref{fig:uncontrolled_new} illustrates the state trajectories diverging to infinity, highlighting the system's instability without control, with feedback gains \(K_1 = K_2 = 0\). We have simulated the open-loop system to illustrate the point that, without control, it is not possible to regulate our nonlinear system trajectories toward zero.

A comparative overview of the three techniques is presented in Table~\ref{tab:comparison_new}, which shows the numerical calculation of quadratic cost in MATLAB and the simulation time.

\begin{table}[ht]
    \centering
    \caption{Comparison of costs and simulation times for the nonlinear system.}
    \vspace{5pt}
    \begin{tabular}{|l|c|c|}
        \hline
        Approach & Cost & Simulation Time (s) \\
        \hline
        Conventional SDRE & 9.2113 & 0.0988 \\
        IRL-based SDRE & 9.2268 & 1.1676 \\
        Uncontrolled & $\infty$ & 0.0079 \\
        \hline
    \end{tabular}
    \label{tab:comparison_new}
\end{table}

As shown in Figures~\ref{fig:sdre_new} and~\ref{fig:irlsdre_new}, both SDRE and IRL-based SDRE effectively regulate the states to the origin, with adaptive gains adjusting to the state-dependent nonlinearity. Table~\ref{tab:comparison_new} indicates that the IRL-based SDRE achieves approximately the same cost as the conventional SDRE (9.2268 vs. 9.2113), and increasing the number of IRL iterations would likely yield identical results. However, the IRL-based SDRE has a simulation time approximately 11 times higher (1.1676 s vs. 0.0988 s) due to its computationally intensive iterative process. This trade-off is acceptable given the partially model-free nature of IRL, which is advantageous when the system model is uncertain. The system without control, depicted in Figure~\ref{fig:uncontrolled_new}, fails to regulate, with states diverging and a cost that converges to infinity.

\section{Conclusion and Future Work}
\label{sec:conclusion}
This study integrates IRL into the conventional SDRE framework, resulting in a new control scheme referred to as the IRL-based SDRE. In this method, instead of solving ARE in every state using linear algebraic toolboxes, we run multiple iterations of IRL, which is partially model-free. Although this process has an advantage over classical model-based methods in terms of model dependency, because IRL iterations demand a high computational load, it needs a lot of computational capacity for real-time simulation.

For those who want to work in this field, there is another variant of IRL for the LQR problem, which is completely model-free, that can be used as an alternative to the IRL used in this article. 
The second idea is that running this in real-time is not possible, and there may be ideas that can increase the speed; a significant amount of research can be conducted for this purpose. Further work is underway to optimize and accelerate the computation process for both the traditional SDRE and the IRL-based SDRE. Linear algebraic theorems can be used to accommodate less frequent changes in control gain while still providing a stability proof.

\section{Code Availability}
The code used in this study has been deposited in a public GitHub repository and is available at:\\
\url{https://github.com/ariarsh/IRL-Based-SDRE}.

\bibliographystyle{IEEEtran}
\bibliography{mybib}

\end{document}